\PassOptionsToPackage{dvipsnames,table}{xcolor}

\documentclass[sigconf, screen]{acmart}

\usepackage[normalem]{ulem}
\usepackage{booktabs}
\usepackage{longtable}
\usepackage{tabularx}
\usepackage{enumitem}
\usepackage{seqsplit}
\usepackage{multirow}
\usepackage{xspace}
\usepackage{array}
\usepackage{graphicx}
\usepackage{amsmath}
\usepackage{tikz}
\usetikzlibrary{arrows.meta, positioning}
\usepackage{listings}
\usepackage{float}
\newfloat{listing}{tbp}{lol}
\floatname{listing}{Listing}

\lstdefinelanguage{WebMazeJSON}{
 morestring=[b]",
 morekeywords={true,false,null},
 sensitive=true
}
\usepackage[ruled,vlined]{algorithm2e}

\usepackage[labelfont=bf,font=small, compatibility=false, skip=5pt]{caption} %

\newcommand{\squishlist}{\begin{itemize}[leftmargin=*,noitemsep,topsep=0pt,parsep=0pt,partopsep=0pt]}
\newcommand{\squishend}{\end{itemize}}

\newcommand{\x}{$\times$\xspace}

\newcommand{\PP}[1]{%
  \begingroup
    \edef\temp{\detokenize{#1}}%
    \IfEndWith{\temp}{.}{%
      \noindent\textbf{#1}%
    }{%
      \noindent\textbf{#1.}%
    }%
  \endgroup
}

\def\Snospace~{\S{}}

\usepackage{pifont}
\newcommand{\WC}[1]{\ding{\numexpr171+#1\relax}}

\begin{document}

\title{WebMaze: Defending Against LLM-Powered Web Scanners with Functionality-Coupled Deception}


\author{Pengfei Zeng}
\affiliation{
\institution{Wuhan University}
\city{Wuhan}
\country{China}
}
\email{pfzeng@whu.edu.cn}

\author{Changhua Luo}
\affiliation{
\institution{Wuhan University}
\city{Wuhan}
\country{China}
}
\email{chluo2502@whu.edu.cn}

\author{Penghui Li}
\affiliation{
\institution{Nanyang Technological University}
\country{Singapore}
}
\email{penghui.li@ntu.edu.sg}

\newcommand{\sys}{\textsc{WebMaze}\xspace}

\settopmatter{printfolios=true}
\begin{abstract}
Large language models (LLMs) have enabled more effective vulnerability scanning of web applications.
However, these capabilities also make it significantly easier for attackers to automate cyberattacks on the web.
In this work, we investigate how to divert malicious LLM-based web scanning, with preservation of genuine functionality as a deployment requirement.
We systematically study six LLM-based scanners and summarize how they gather information and iteratively select actions based on observations.
Guided by the study, we then present \sys, a defense that preserves genuine application functionality while diverting LLM-based scanners to a generated decoy world.
\sys embeds functionality-coupled traps in genuine pages and constructs view-tailored interactive decoy worlds from the application's existing views.
The traps blend into genuine components and expose entry points to decoy worlds that sustain exploration through connected pages and interactions without changing genuine functionality.
Across 12 web applications and three LLM-based scanners under
the default no-disclosure setting, decoy pages account for an
average of 93.0\% of the distinct pages requested per run.
This demonstrates that \sys effectively diverts scanners away from genuine application pages and into the generated decoy worlds.
Even when explicitly informed about the decoy mechanism and instructed to avoid decoy pages, scanners continue to explore predominantly decoy pages.
\end{abstract}

\maketitle

\section{Introduction}
\label{sec:introduction}

Web scanning is an effective approach for discovering security vulnerabilities in web applications.
Recent advances in large language models (LLMs) have further enhanced its capabilities by integrating LLMs with browsers, direct HTTP requests, and security testing tools.
Unlike traditional scanners that primarily rely on predefined rules or exploration strategies, LLM-based scanners can reason about application states, determine subsequent actions, and dynamically adapt their strategies based on observations~\cite{deng2024pentestgpt,stafeev2025yura}.
These capabilities have enabled the discovery of security vulnerabilities in widely deployed web applications, including WordPress~\cite{kues2026wp2shell}.

Despite their benefits for legitimate security testing, LLM-based scanners can also be exploited for malicious purposes.
By substantially reducing the expertise and effort required to discover and exploit vulnerabilities, they lower the barrier to conducting sophisticated web attacks.
Consequently, attackers have begun leveraging LLM-powered agents to automate and scale their attacks~\cite{anthropic2025espionage, kues2026wp2shell}.

There is an urgent need to protect web applications against the growing threat posed by adversarial LLM-based web scanners.
Existing defenses are not sufficient.
One category of work tries to distinguish legitimate user requests from abnormal scanning via request-pattern detection~\cite{azad2020webrunner,wan2019pathmarker, searles2023captchas}.
However, they have been demonstrated to be easily bypassed or evaded~\cite{deng2025oedipus, teoh2025captchas}.
They also impose friction on legitimate users~\cite{searles2023captchas}.
Another line of work employs deception to mislead scanners and divert their exploration toward decoys.
Prior work has explored techniques to disrupt or divert automated exploration, including URL marking, honeypot replicas, and misleading webpage content~\cite{wan2019pathmarker,tsouvalas2025parallax,li2026webcloak}.
However, deception still faces two challenges.
First, 
deception uses \emph{traps}, which are entry points embedded in genuine webpages to divert automated scanners toward decoy functionality.
Scanners can identify and filter traps once they know their patterns.
For example, hidden-link traps use visually concealed links to decoy pages, which scanners can easily filter by excluding links that are not visibly rendered.
Second, even when scanners enter decoy environments, limited functionality and interaction opportunities may cause them to abandon further exploration.  Related work on LLM-powered honeypots has independently noted analogous limitations, including insufficient realism, limited adaptability and shallow interaction~\cite{sladic2024honeypot}.
These challenges motivate two requirements for effective deception: (1) decoy entry points must be difficult to identify and filter, and (2) decoy environments must sustain continued scanner exploration.

To understand how to meet these requirements, we study six state-of-the-art LLM-based web scanners and examine how they explore web applications.
We identify a recurring three-stage \emph{perception--decision--feedback} loop (\autoref{sec:scanner-study}).

Guided by the study, our key insight is that deception becomes more robust when it is coupled to genuine application functionality.
If a deceptive element uses a pattern that is absent or different from genuine functionality, a defense-aware scanner can identify and reject it at relatively low cost.
In contrast, if the same pattern also appears in legitimate components, rejecting all instances of that pattern may cause the scanner to miss genuine functionality.
The scanner must then either accept some decoy interactions or invest additional effort in separating them from legitimate ones.
We refer to this tension as the \emph{evasion-coverage tradeoff}.
Moreover, once a scanner enters a decoy, the decoy must provide sufficiently rich and plausible functionality to sustain exploration and consume the scanner's limited budget.

We present \sys, a framework that employs functionality-coupled deception to divert LLM-based web scanners away from genuine application functionality.
\sys consists of two components.
First, \sys embeds functionality-coupled traps into genuine application pages.
These deceptive entry points \emph{reuse} the structures and styles of web components already present in the site's legitimate workflows. %
For example, some websites use small elements near page boundaries to trigger the preloading of additional content.
\sys adopts a similar design by placing a one-pixel decoy link near the page boundary, directing scanners that follow it into a decoy world.
Its small size and peripheral placement make legitimate users unlikely to click it during normal browsing.

Second, \sys constructs view-tailored decoy worlds that mimic the protected application's views and interaction conventions.
These worlds connect generated pages through decoy routes and provide pre-defined responses to supported interactions.
This allows scanners to continue navigating and interacting as if they were exploring genuine application functionality.
Decoy handlers use designated resources and state stores; genuine request handling must satisfy the preservation conditions.

We evaluate \sys on 12 web applications using three LLM-based scanners and two model backbones.
Across all experiment runs, decoy pages account for an average of 93.0\% of the distinct pages requested by the scanners.
Even under the most informative disclosure condition, where scanners are informed of the decoy strategies, the mean decoy-page share among tested pages ranges from 85.6\% to 94.8\%.
This demonstrates the effectiveness of \sys in diverting web scanners from vulnerability scanning.

In summary, this work makes the following contributions:
\squishlist
    \item We study six LLM-based web scanners and identify a shared three-stage loop in their scanning mechanisms.
    \item We introduce functionality-coupled deception and design \sys to combine traps embedded in genuine pages with interactive, view-tailored decoy worlds.
    \item We demonstrated \sys's effectiveness across 12 web applications and three scanners.
    \item We will release \sys as open source soon.
\squishend

\section{Background}
\label{sec:background}

\subsection{Web Scanning}
Web vulnerability scanning aims to identify security weaknesses by automatically exploring web applications and testing their exposed interfaces~\cite{fong2007web}.
Traditional scanners crawl applications by following links, identifying forms, and sending crafted requests to discovered entry points.
However, they often struggle to reach deeper application states that require multi-step workflows~\cite{eriksson2021blackwidow}.

Recent approaches incorporate LLMs to address these limitations by interpreting webpage content and planning actions based on application context~\cite{deng2024pentestgpt,stafeev2025yura}.
LLMs enable scanners to reason about application functionality and execute sequences of interactions similar to those of human users.
For example, a scanner may identify a login form, submit valid credentials, navigate to authenticated pages, and explore functionality available after login.
LLM-based scanners can also invoke external tools, such as security testing utilities, to execute actions and use their results to guide subsequent exploration and exploitation.

\subsection{Existing Defenses Against Web Scanning}
To protect web applications against malicious automated scanning, 
existing defenses commonly rely on request-pattern detection, CAPTCHAs, and deception.
Request-pattern detection analyzes access frequency, navigation paths, and timing to identify and block suspicious clients~\cite{azad2020webrunner,wan2019pathmarker}.
CAPTCHAs require users to complete verification challenges intended to distinguish humans from automated clients~\cite{searles2023captchas}.
However, these defenses operate during normal site use and may disrupt legitimate users.
For instance, detection incorrectly blocks benign clients, while CAPTCHA challenges introduce additional friction and can be solved by automated systems.

Deception offers an alternative defense strategy by misleading automated scanners and diverting their exploration toward decoys, while minimizing interference with legitimate users~\cite{ayzenshteyn2025cheat,mueter2008honeypots,tsouvalas2025parallax,li2026webcloak}.
For instance, CHeaT embeds misleading information, such as fabricated vulnerability reports, in HTML comments to divert LLM-based agents~\cite{ayzenshteyn2025cheat}.
However, such defenses can be bypassed when scanners are aware of their strategies.
For instance, a scanner aware of CHeaT's comment-based traps could simply remove HTML comments before analyzing webpage content.

\subsection{Existing LLM-Based Scanners}
\label{sec:scanner-study}
We analyze how LLM-based scanners explore web applications to identify requirements for effective deception.
Instead of relying on ad hoc traps that can be easily bypassed, these requirements should guide the design of defenses that can withstand adaptive scanners.

\noindent\textbf{Study methodology.}
We systematically examine six publicly available LLM-based scanning and penetration-testing systems, covering browser-based, HTTP-based, multi-agent, and human-in-the-loop approaches.
They are YuraScanner~\cite{stafeev2025yura}, HackingBuddyGPT~\cite{happe2023hackingbuddygpt}, AWE~\cite{jaswal2026awe}, PentestGPT~\cite{deng2024pentestgpt}, Cochise~\cite{happe2026cochise}, and PentestAgent~\cite{shen2025pentestagent}.
\autoref{tab:scanners} summarizes the scanners we examined.

\noindent\textbf{Three-stage loop.}
We observe that the six scanners follow a recurring three-stage exploration loop: \emph{perception}, \emph{decision}, and \emph{feedback}.
\emph{Perception} gathers information about the application, such as webpage content and HTTP responses.
\emph{Decision} combines these observations with accumulated agent-side context, including previously executed actions and testing results, to select and execute the next action.
Common actions include visiting pages, submitting forms, and testing vulnerabilities with attack payloads.
\emph{Feedback} incorporates the exploration results into this agent-side context.
In the next cycle, the scanner perceives the resulting webpage or response and uses this information to determine its next action.

\begin{table}[!t]
\centering
\scriptsize
\caption{Studied scanners, their approaches, and exploration loops.}
\label{tab:scanners}
\renewcommand{\arraystretch}{1.12}
\resizebox{\columnwidth}{!}{%
\begin{tabular}{@{}lllll@{}}
\toprule
Agent & Approach & \multicolumn{3}{c}{Three-stage Exploration Loop} \\
\cmidrule(lr){3-5}
 & & Perception & Decision & Feedback \\
\midrule
YuraScanner
    & Browser-based
    & DOM and page state
    & Browser action
    & Refresh page state \\
HBGPT
    & HTTP-based
    & HTTP response
    & HTTP command
    & Store response \\
AWE
    & Multi-agent
    & Page and test results
    & Endpoint or payload
    & Update verdict \\
PentestGPT
    & Human-in-the-loop
    & Tool output
    & Testing subtask
    & Update task tree \\
Cochise
    & Tool-based
    & Shell output
    & Testing task
    & Update knowledge \\
PentestAgent
    & Multi-agent
    & Reconnaissance results
    & Vulnerability test
    & Update plan \\
\bottomrule
\end{tabular}%
}
\end{table}

\noindent\textbf{Requirements for deception}
Based on this common exploration loop, we derive two requirements for effective deception.
\emph{First, deceptive entry points must be difficult to identify and filter.}
All six scanners rely on information obtained from the target application to guide their exploration.
Deceptive content can therefore influence scanners' subsequent actions.
However, defense-aware scanners may identify and filter deceptive entries.
Thus, deceptive entry points should be closely coupled with genuine functionality, making them difficult to filter without sacrificing legitimate exploration.

\emph{Second, decoy environments must sustain continued exploration.}
All six scanners update this accumulated agent-side context after executing an action and use the updated information to guide subsequent decisions.
Consequently, merely attracting scanners into a decoy is insufficient if they quickly encounter dead ends.
Decoy environments should provide connected pages and interactions that allow scanners to continue their exploration.

\subsection{Threat model and research goals.}
We aim to divert LLM-based scanners toward decoy functionality without disrupting legitimate workflows, subject to the compatibility conditions described below.
We consider a website operator (or defender) who controls the protected application.
The defender can modify application pages and deploy decoy functionality but cannot control the scanner's internal reasoning, prompts, or tools.

The attacker employs an LLM-based scanner to discover and exploit vulnerabilities.
We assume that the attacker knows the existence and mechanisms of the defense and can instruct the scanner to avoid decoys.
The scanner operates under finite budgets for requests, execution time, and LLM interactions.

\section{Methodology}
\label{sec:methodology}

\begin{figure*}[t]
\centering
\includegraphics[width=\textwidth]{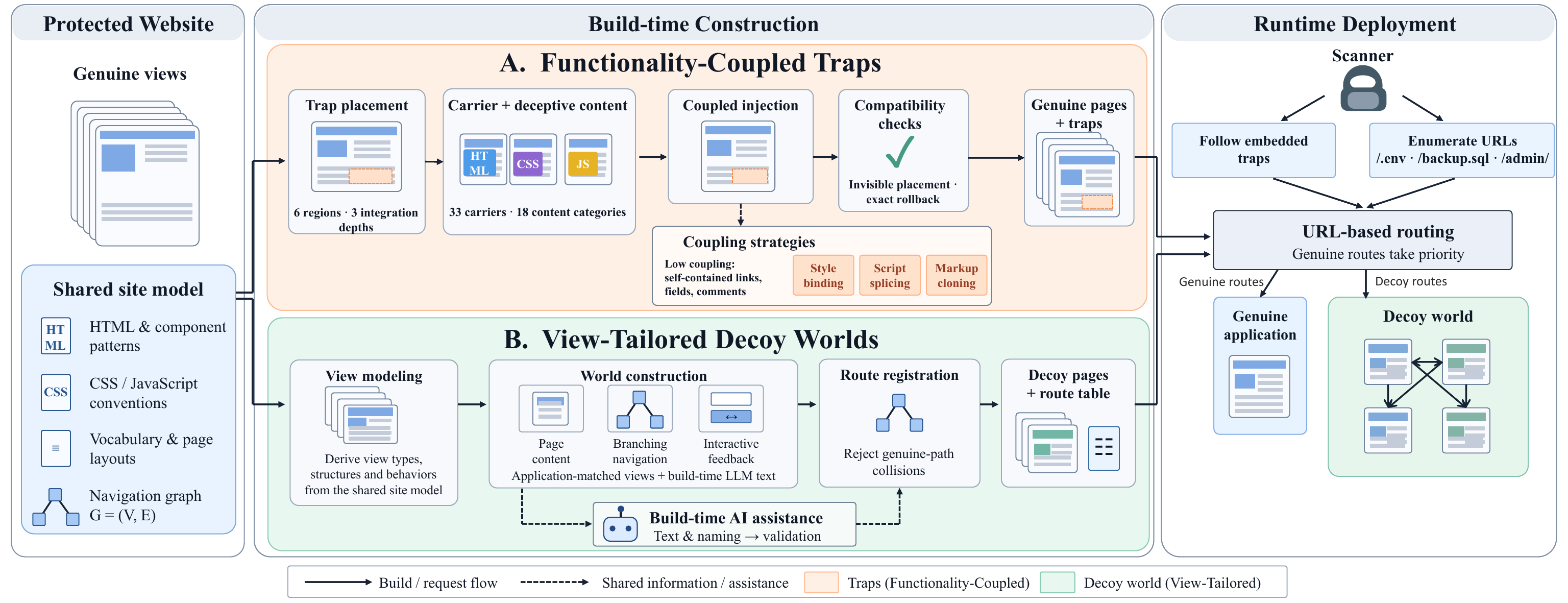}
\Description{The protected site's views feed build-time trap and decoy-world construction; runtime routing separates genuine and decoy interactions.}
\caption{\sys architecture: functionality-coupled traps and view-tailored decoy worlds, generated at build time and routed at run time.}
\label{fig:system}
\end{figure*}

We propose \sys, a framework that addresses these two requirements through functionality-coupled traps and a view-tailored decoy world.
Functionality-coupled traps reuse the structures and interaction patterns of genuine application components and direct scanners into the decoy world.
This makes the traps difficult to identify and filter without also excluding legitimate functionality.
View-tailored decoy worlds provide connected pages and interactions similar to those of the protected application, making it difficult for scanners to determine that they have entered a decoy world.
\autoref{fig:system} illustrates how \sys constructs and deploys these components to divert scanner exploration.

\subsection{Functionality-Coupled Traps}
\label{sec:traps}

A defense-aware scanner can filter out easily identifiable standalone traps.
Functionality-coupled traps instead reuse the structures and interaction patterns of genuine application components, making broad filtering likely to exclude legitimate functionality.

\subsubsection{Trap Construction}
\label{sec:method-trap-construction}

\sys constructs traps from templates and customizes them for
the target web applications.

\begin{listing}[t]
\begin{lstlisting}[language=HTML, frame=single]
<!-- (a) Trap template -->
<link rel="preconnect" href="{{fake_target}}"
      style="{{zero_size_style}}">

<!-- (b) Application context -->
<head>
  <title>Login :: Damn Vulnerable Web Application
    (DVWA) v1.10 *Development*</title>
  <link rel="stylesheet" type="text/css"
        href="dvwa/css/login.css" />
  !!!Trap in`sertion point here.
</head>

<!-- (c) Instantiated trap -->
<link rel="preconnect" href="/api/v1/settings/9"
      style="width:calc(0px);height:calc(0px);
             padding:0;border:0;margin:0">
\end{lstlisting}
    \caption{Constructing a DVWA trap: (a) a template,
    (b) abbreviated source context, and (c) an instantiated trap.}
    \label{code:template-example}
\end{listing}

\noindent\textbf{Trap template.}
We define each trap template in \sys along three axes.
A \emph{carrier type} defines the HTML or script mechanism used to embed a trap, a \emph{content category} defines the resource or information it suggests, and a \emph{visibility technique} defines how the trap is rendered.
For example, links to \texttt{/admin/login} and \texttt{/admin/dashboard} count as two trap instances, but both use the same carrier type (HTML link) and content category (administrative path).
Replacing a link with an HTML comment changes the carrier type, while replacing an administrative URL with a configuration-file URL changes the content category.
Similarly, replacing zero-size styling with off-screen positioning changes the visibility technique.
Strategies for instantiating templates affect individual trap instances but must conform to the template's carrier type, content category, and visibility technique.

\noindent\textbf{Application-aware trap generation.}
\sys instantiates trap templates to generate traps tailored to the protected application.
It first analyzes the website to identify existing components, their locations, and naming conventions.
For each template, \sys then finds components that match its \emph{carrier} type and selects nearby locations where such carriers already appear.
It generates deceptive \emph{content} consistent with the application's conventions and applies the template's \emph{visibility} technique.
For example, suppose an administration page contains navigation links with the CSS class \texttt{nav-link}, using paths such as \texttt{/admin/users}, and also includes a hidden HTML link.
\sys can place the trap alongside the existing navigation links, reuse the \texttt{nav-link} class and the page's existing hidden-link presentation, and generate a decoy path such as \texttt{/admin/decoy}.

\noindent\textbf{DVWA example.}
We use DVWA to illustrate how \sys instantiates a trap for a protected web application.
\autoref{code:template-example}(a) shows a template whose \emph{carrier} is an HTML resource hint defined by \texttt{rel="preconnect"}.
The \emph{content} is specified through the \texttt{href} attribute, while the \emph{visibility} is specified through the \texttt{style} attribute.
As shown in (b), DVWA's login page already contains a stylesheet declaration in its HTML head.
\sys follows this structure by inserting another \texttt{link} element before the closing head tag.
The instantiated trap in (c) points to the decoy route \texttt{/api/v1/settings/9} and applies zero-size styling.

\subsubsection{Trap Integration}
\label{trap-inte}

After constructing traps, \sys analyzes the protected application to identify existing patterns suitable for their integration.
It then adapts each trap to the surrounding page without interfering with normal browsing.

\PP{Low-Coupling Integration}
Low-coupling traps are added as independent components without modifying existing application components.
\sys places them at locations where similar links, forms, or actions already appear and adapts their text, attributes, and placement to the surrounding page.
Because they neither replace genuine components nor participate in the application's original execution paths, they preserve existing functionality.
Their visibility techniques further make accidental interaction by legitimate users unlikely, preserving the normal browsing experience.
These properties allow low-coupling traps to be deployed broadly across pages with little modification to the application.

\PP{High-Coupling Integration}
High-coupling traps reuse or extend existing application structures so that separating them from genuine components requires analyzing their relationship with the surrounding code.
\sys realizes this integration through style binding, script splicing, and markup cloning.

\emph{Style Binding.}
\sys identifies existing CSS rules whose styling can be reused by traps.
It extends their selectors with trap-specific classes that follow the application's naming conventions.
For example, if a webpage uses \texttt{.hidden-form} to style forms that are not displayed by default, \sys can extend the same rule to \texttt{.hidden-form-alt} and assign this class to a decoy form.
The decoy form therefore shares the styling of genuine components while remaining unlikely to be accessed accidentally by legitimate users because it is not visibly rendered during normal browsing.

\emph{Script Splicing.}
\sys identifies locations in existing JavaScript where deceptive statements can be inserted without changing the application's original logic.
Rather than modifying identifiers or adding misleading comments, \sys inserts valid statements that follow the surrounding code structure and naming conventions.
For example, if a script declares \texttt{apiEndpoint}, \sys can insert an independent declaration \texttt{apiEndpointBackup} that refers to a decoy URL.
Because the new variable is not referenced by the original code, existing execution paths remain unchanged, while scanners inspecting the script may discover and follow the decoy URL.

\emph{Markup Cloning.}
\sys identifies existing HTML structures that can be reused to construct traps.
The generated traps follow the markup and placement conventions of genuine components without replacing them.
For example, if a webpage places small auxiliary links near page boundaries for content preloading, \sys can reuse the same markup structure for a one-pixel decoy link in the same region.
The trap therefore follows the application's existing structure while remaining unlikely to receive accidental clicks from legitimate users.

\subsection{View-Tailored Decoy Worlds}
\label{sec:fake-world}

A deceptive entry point alone is insufficient to sustain scanner exploration.
Even after entering a decoy, LLM-based scanners may abandon it if interactions fail or the pages differ substantially from the protected application.
For example, a nonfunctional form or a school management page appearing within a shopping website may reveal inconsistencies that cause scanners to return to genuine functionality.
To address this challenge, \sys automatically constructs view-tailored decoy worlds that mimic the protected application's page structures and interaction patterns.
These worlds support continued exploration while keeping decoy interactions isolated from the genuine application's functionality and data.

\subsubsection{View Modeling}

To make decoy pages consistent with the protected application, \sys first constructs a view model from its existing pages.

\sys captures the HTML documents of the application's existing routes.
Each captured document serves as a source view that can be reused to generate decoy pages.
For each source view, \sys records its original route and extracts its page structure.
A captured document represents one observed page state and does not necessarily cover every state of the corresponding route.

\sys also extracts conventions shared across the application's pages.
It analyzes existing CSS rules to determine how the application styles its components.
Meanwhile, it extracts links between captured pages to construct a navigation graph that describes the application's existing navigation structure.

The resulting view model provides reusable page structures and application-specific conventions for decoy-world construction.
It captures the application's presentation and navigation without reproducing its underlying business logic.

\subsubsection{Decoy World Construction}

The view model guides \sys in constructing a decoy world that follows the protected application's page structures and navigation patterns.
Based on this model, \sys generates connected decoy pages and provides responses to supported interactions, allowing scanners to continue exploring within the decoy world.
\autoref{alg:fake-world} summarizes the construction process.

\begin{algorithm}[t]
    \caption{Construction of a view-tailored decoy world.}
    \footnotesize
    \label{alg:fake-world}
    \SetKwInOut{Input}{Input}
    \SetKwInOut{Output}{Output}
    \SetKwFunction{BuildSeed}{BuildSeed}
    \SetKwFunction{Plan}{PlanRoutesAndGraph}
    \SetKwFunction{Customize}{CustomizeContent}
    \SetKwFunction{Render}{RenderPage}
    \SetKwFunction{Bind}{BindImplementedHandlers}
    \Input{View model $M$, deployment secret $k$, site identifier $n$, construction budget $B$}
    \Output{Decoy world $W_D=(V_D,E(G_D),H_D)$ and decoy-route map $R_D$}

    $\sigma \leftarrow$ \BuildSeed{$k,n$}\;
    $(P,G_D) \leftarrow$ \Plan{$M,\sigma,B$}\;
    $\mathcal{D} \leftarrow$ \Customize{$M,P$}\;
    $V_D \leftarrow \emptyset$\;

    \ForEach{$d \in V(G_D)$}{
        $s \leftarrow P[d].\mathrm{template}$\;
        $v_d \leftarrow$ \Render{$M.\mathrm{source}[s],M.\mathrm{style},d,\mathcal{D}[d],P$}\;
        $V_D \leftarrow V_D \cup \{v_d\}$\;
    }

    $F_D \leftarrow \textsc{MaterializeResources}(M,P,\mathcal{D},V_D)$\;
    $R_D \leftarrow \textsc{RegisterFakeResources}(P,V_D,F_D)$\;
    $H_D \leftarrow$ \Bind{$R_D$}\;

    \Return{$W_D=(V_D,E(G_D),H_D),R_D$}\;
\end{algorithm}

\noindent\textbf{Navigation Planning.}
To provide continued exploration opportunities, \sys plans a directed graph of decoy pages before generating their content.
Unlike the navigation graph extracted from the protected application, this graph connects newly generated decoy routes.
The planned routes follow the application's URL conventions and include links between related pages.
Each route is assigned a source view from the view model, allowing subsequent pages to retain the application's structure.

\noindent\textbf{Page Instantiation.}
After planning the navigation graph, \sys generates each decoy page from its assigned source view.
It reuses the original HTML structure and styling conventions while adapting the content to the planned decoy routes.
This allows newly generated pages to resemble the protected application while providing additional exploration opportunities.

An LLM further customizes the generated pages by producing content consistent with the protected application's domain and terminology.
For example, for a shopping website, the LLM can generate product descriptions that resemble those on genuine pages.
\sys validates the generated content before incorporating it into the decoy pages.
This increases the diversity of decoy pages even when only a limited number of source views are available.

\noindent\textbf{Interaction.}
Navigation alone is insufficient if scanners cannot interact with the generated pages.
Therefore, \sys assigns request handlers to decoy resources to support further interactions.
These handlers return HTTP responses or generate additional decoy pages.
However, they would not modify the protected application's backend or data (see \autoref{iso} and \autoref{sec:routing}).
Specifically, \sys implements three types of request handlers.

\WC{1} \emph{Static handlers} return stored resources, such as HTML pages and JSON files.
For example, submitting a form whose destination is a static decoy page returns the corresponding page.

\WC{2} \emph{Parameterized handlers} generate responses for routes containing dynamic parameters.
They use prepared content to produce deterministic responses.
This allows scanners to explore additional routes without requiring every possible page to be generated in advance.

\WC{3} \emph{Stateful handlers} support interactions that require persistent data.
\sys maintains a separate decoy database and implements supported read and write operations.
For example, creating an account updates the decoy database, allowing subsequent requests to retrieve the newly created account.
These operations do not affect the protected application's database.
To reduce the attack surface for potentially resource-intensive state-changing operations, such as repeated record creation, \sys limits the number of stateful handlers.
For other routes, \sys pseudorandomly selects between static and parameterized handlers.

\subsubsection{Routing Between the Genuine Application and Decoy World}
\label{sec:routing}

The decoy world introduces additional routes alongside those of the protected application.
Decoy routes must not overlap with existing application routes to prevent legitimate users from accidentally reaching decoy pages by manually entering a URL.
To achieve this, \sys registers decoy routes separately and uses URL-based routing to dispatch requests to the appropriate destination.

\noindent\textbf{Route Registration.}
\sys first identifies the protected application's existing routes.
It then checks each generated decoy route against these routes and excludes conflicting candidates from registration.
For example, if the application already provides \texttt{/admin/}, \sys does not register a decoy at that path.
This prevents known genuine routes from being overwritten.

\noindent\textbf{Request Dispatch.}
After registering the decoy routes, \sys uses the resulting route table to determine whether incoming requests should be handled by the genuine application or the decoy world.
Incoming URL paths are normalized before route matching.
We model route ownership as a partial mapping
\begin{equation}
R:P\rightharpoonup\{\mathrm{gen},\mathrm{decoy}\},
\end{equation}
where $P$ is the set of normalized paths.
Parameterized decoy routes are registered as route patterns.

For each incoming request, \sys first checks whether its URL matches a registered decoy route.
Two types of routes lead to the decoy world.
The first consists of entry routes exposed by traps in genuine pages (\autoref{trap-inte}), such as the destination of a hidden form.
The second consists of internal decoy routes that connect generated pages and support continued exploration.
Requests matching either type are handled by the decoy world, while other requests follow the genuine application's routing.
Thus, \sys diverts scanner exploration without interfering with unmatched genuine requests.

\subsubsection{Isolation}
\label{iso}
In addition to browser experiences,
decoy pages must not affect the protected application's functionality or data.
Separating genuine and decoy routes alone is insufficient.
JavaScript running within decoy pages may still issue requests to genuine APIs, while stateful decoy interactions require persistent storage.
To address these risks, \sys isolates API requests and data storage from the protected application.

\noindent\textbf{API Isolation.}
\sys applies a separate API request policy to generated decoy pages.
Only requests targeting registered decoy API endpoints are permitted.
Requests targeting the protected application's APIs are blocked before reaching the genuine backend.
This restriction applies only to decoy pages, leaving API access from genuine application pages unchanged.

\noindent\textbf{Storage Isolation.}
\sys maintains a separate database for the decoy world.
All stateful decoy operations access this database rather than the protected application's database.
The decoy database is initialized with synthetic data instead of genuine application records.
Consequently, even if a scanner retrieves the entire decoy database, it cannot directly obtain records from the protected application's database.
This separation introduces additional storage overhead.
To keep this overhead manageable, \sys limits resource-intensive decoy operations, such as persistent writes through stateful handlers.
Most decoy requests return pre-generated responses without executing genuine backend operations or modifying persistent data.

\section{Evaluation}
\label{sec:evaluation}

We evaluate \sys to answer the following questions.

\squishlist

\item \textbf{RQ1 (Defense Effectiveness)}: How effective is \sys at diverting scanners from genuine vulnerability targets?

\item \textbf{RQ2 (Ablation Study)}: How do the components in \sys contribute to defense effectiveness?

\item \textbf{RQ3 (Scanner Awareness)}: How robust is \sys when scanners know about the decoy mechanism?

\item \textbf{RQ4 (Statistics)}: What are the size and composition of the decoy worlds generated by \sys across applications?
\squishend

\subsection{Experimental Setup}
\label{sec:eval-setup}

\PP{Target Scanners}
We evaluate three state-of-the-art LLM-based web scanners: AWE~\cite{jaswal2026awe}, hackingBuddyGPT (HBGPT)~\cite{happe2023hackingbuddygpt}, and YuraScanner~\cite{stafeev2025yura}.
We select them because they can be deployed reliably in automated experiments and represent three distinct paradigms of LLM-powered web penetration testing.
Specifically, YuraScanner performs browser-based task-driven scanning, HBGPT conducts tool-driven exploration, and AWE performs module-based security testing.
Other candidate scanners require substantially more cumbersome setup or support narrower application ranges, making large-scale evaluation difficult.
Across all experiments, AWE and YuraScanner are limited to twenty minutes per run, while HBGPT is limited to one hundred rounds.
We evaluate each scanner with two model backbones, DeepSeek V4.1 Flash~\cite{deepseek-v4.1flash} and GLM-5.3~\cite{glm-5.3}, and repeat each scanner--application--backbone combination three times.

\PP{Application Dataset}
Our dataset comprises 12 web applications spanning PHP, Java,
JavaScript/Node.js, Python, and static HTML.
DVWA~\cite{dvwa}, Mutillidae~\cite{owasp-mutillidae-ii}, and the OWASP Benchmark suite~\cite{owasp-benchmark} contain the ground truth and we use them to evaluate vulnerability detection capabilities of scanners. 
Specifically, DVWA and OWASP Benchmark provide labeled vulnerability-reference URLs, whereas Mutillidae provides vulnerability-triggering pages.
We obtain the remaining 9 applications from prior web scanning work~\cite{eriksson2021blackwidow, witcher}.
For these applications, we do not have complete vulnerability ground truth because establishing it would require validating every route, parameter, and application state, which is difficult to do exhaustively.
We include them to cover different application scenarios and to assess if \sys is applicable to different web techniques.

\PP{Metrics}
We use two coverage metrics to evaluate scanner exploration and defense effectiveness.

\emph{Vulnerability-Page Coverage}
For the three applications with ground-truth vulnerabilities, we measure the fraction of distinct vulnerability-triggering pages reached by each scanner.
Because the evaluated scanners do not support all vulnerability types in our dataset, we do not require them to successfully trigger each vulnerability.
Reaching the corresponding page is a prerequisite for testing a vulnerability and therefore provides a proxy for vulnerability detection coverage.

\emph{Decoy-Page Coverage}
We measure how effectively \sys redirects scanner exploration using decoy-page coverage, defined as the fraction of distinct requested URLs that belong to the decoy world:
\[
\mathrm{Coverage}
=
\frac{\#\text{ distinct decoy URLs}}
{\#\text{ distinct requested URLs}}.
\]
Higher coverage means that a larger fraction of scanner exploration is directed to the decoy world, indicating stronger defense effectiveness.

\subsection{RQ1: Defense Effectiveness}
\label{sec:eval-main}

\subsubsection{Vulnerability-page Coverage}
\label{sec:eval-url-contact}
We evaluate whether \sys reduces scanners' access to
vulnerability-related pages.
\autoref{tab:rq1-vulnerability} reports the fraction of distinct vulnerability-triggering pages reached by each scanner.
Without a defense, all three scanners reach 100\% of vulnerability-triggering pages on DVWA.
Their coverage on OWASP ranges from 73.9\% to 100\%, while their Mutillidae coverage ranges from 11.8\% to 47.1\%.
These results show that the scanners can reach most vulnerability-triggering pages in the original applications, with coverage varying across scanners and applications.
Note that the two model backbones produced similar vulnerability-page coverage, so we report the DeepSeek V4.1 Flash results here.
We next present defense effectiveness.

\PP{\sys}
With \sys, coverage drops substantially.
All three scanners reach none of the DVWA vulnerability-triggering pages and only 4.41\%--4.90\% of the Mutillidae candidate pages.
On OWASP, AWE and HBGPT reach none of the vulnerability-triggering pages, while YuraScanner reaches only 1.00\%.
Compared with the original applications, \sys substantially reduces scanners' access to vulnerability-triggering pages.
Because reaching a vulnerability-triggering page is a prerequisite for testing the corresponding vulnerability, this reduction limits the scanners' opportunity for vulnerability detection.

\PP{CHeat~\cite{ayzenshteyn2025cheat}}
CHeat is a relevant work that inserts misleading instructions into HTML pages to influence scanner decisions.
We include it as a comparison of a defense that changes scanner input without constructing a connected decoy world.
As shown in \autoref{tab:rq1-vulnerability}, CHeat reduces vulnerability-page coverage for some scanner--application pairs, but does not do so consistently.
On DVWA, CHeat coverage remains 91.7\% for AWE and HBGPT and 100\% for YuraScanner.
On Mutillidae, coverage ranges from 10.3\% to 25.0\%, and on OWASP it ranges from 11.4\% to 100\%.
In contrast, \sys consistently reduces vulnerability-page coverage across all scanners and applications.

\begin{table}[t]
\centering

\caption{Vulnerability-page coverage under the original applications, CHeat, and \sys using the DeepSeek V4.1 Flash backbone.
Values are normalized by the total number of vulnerability pages.}
\label{tab:rq1-vulnerability}
\resizebox{\columnwidth}{!}{%
\setlength{\tabcolsep}{2.5pt}
\begin{tabular}{@{}l rrr rrr rrr @{}}
\toprule
 & \multicolumn{3}{c}{Original} &  \multicolumn{3}{c}{With \sys}  & \multicolumn{3}{c}{With CHeat} \\
\cmidrule(lr){2-4}\cmidrule(l){5-7}\cmidrule(l){8-10}
Scanner & DVWA & Mutillidae & OWASP & DVWA & Mutillidae & OWASP & DVWA & Mutillidae & OWASP \\
\midrule
AWE         & 100\% & 11.8\% & 100\% & 0.00\% & 4.41\% & 0.00\% & 91.7\% & 10.3\% & 100\%  \\
HBGPT       & 100\% & 47.1\% & 100\%  & 0.00\% & 4.90\% & 0.00\% & 91.7\% & 19.1\% & 11.4\% \\
YuraScanner & 100\% & 11.8\% & 73.9\% & 0.00\% & 4.41\% & 1.00\% & 100\% & 25.0\% & 75.0\% \\
\bottomrule
\end{tabular}
}
\end{table}

\subsubsection{Decoy-Page Coverage}
\label{sec:eval-decoy-coverage}
We first evaluate whether \sys diverts scanners toward the decoy world.
As summarized in \autoref{tab:rq1-summary}, across 216 runs (12 apps \x 3 scanners \x 2 backbones \x 3 repetitions), 
coverage remains high for all three scanners under both backbones, ranging from 88.9\% to 97.3\%.
Each URL is counted once within a run, so repeated requests cannot inflate this measure.
The mean decoy-page coverage is 93.0\% across all runs; and 197 runs reach at least 80\% coverage.
Changing the backbone changes coverage slightly across the scanners. 
AWE and YuraScanner obtain similar mean coverage under the two
backbones, while HBGPT decreases from 97.3\% to 94.2\%.
This indicates that the choice of model has little influence on the defense performance of \sys under the evaluated conditions.

There are some performance differences among scanners, particularly between HBGPT, AWE and YuraScanner.
Our manual analysis found that scanner interaction interfaces and exploration strategies have the most of influence on the results.
Specifically, AWE and HBGPT enumerate paths using a human-predefined dictionary, which can possibly reach decoy routes even without in-page traps.
However, browser-based YuraScanner only follows visible DOM links.
As a result, it is harder to discover decoy routes.
Nevertheless, these results still suggest that \sys diverts a substantial fraction of scanner exploration to the decoy world and that this effect is not tied to the backbone or scanner.

\begin{table}[t]
\centering\small
\caption{Decoy-page coverage across scanner backbones. 
}
\label{tab:rq1-summary}
\setlength{\tabcolsep}{5pt}
\begin{tabular}{@{}lrr@{}}
\toprule
Scanner & DeepSeek V4.1 Flash & GLM-5.3 \\
\midrule
AWE & 94.3\% & 94.3\% \\
HBGPT & 97.3\% & 94.2\% \\
YuraScanner & 88.9\% & 88.9\% \\
\bottomrule
\end{tabular}
\end{table}

\begin{figure}[t]
\centering
\includegraphics[width=\columnwidth]{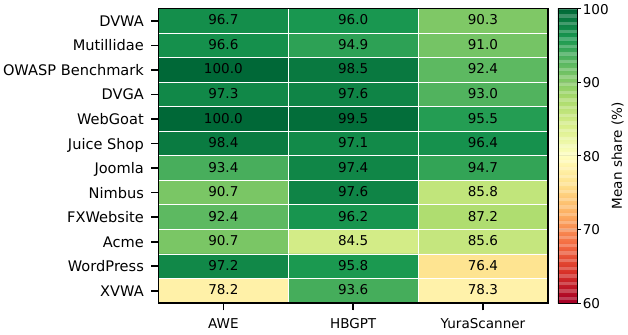}
\Description{Heat map with twelve application rows and three
scanner columns. Each cell reports mean decoy-page coverage
under no disclosure, averaged over two backbones and three
repetitions.}
\caption{Per-application decoy-page coverage under the default
no-disclosure setting. Each cell averages six runs across
two backbones and three repetitions.}
\label{fig:rq1-target-heatmap}
\end{figure}

The per-application breakdown in \autoref{fig:rq1-target-heatmap}
shows coverage of at least 90\% in 29 of the 36
scanner--application cells.
AWE and HBGPT reach this level on 11 of the 12 applications,
and YuraScanner on seven.
The lowest cell means are YuraScanner on WordPress (76.4\%),
AWE on XVWA (78.2\%), and YuraScanner on XVWA (78.3\%).
The remaining lower values are concentrated in a few scanner--application pairs rather than across an entire application family.
Thus, \sys's generated decoy pages occupy most of each scanner's explored page set across diverse applications.

CHeat~\cite{ayzenshteyn2025cheat} directly inserts instructions into original HTML and does not construct a connected decoy world. We thus do not compare \sys with CHeat regarding decoy-page coverage.

\subsection{RQ2: Ablation Study}
\label{sec:eval-ablation}

In the ablation study, we measure how the key components in \sys contribute to its defense effectiveness.
\sys combines low-coupling and high-coupling decoys to provide different entries into the decoy world.
To isolate their roles, we compare the full configuration with variants that remove one or both decoy classes while keeping the decoy-world content and scanner settings unchanged.
\squishlist
\item \textbf{No low-coupling decoys}: removes low-coupling decoys but retains high-coupling decoys.
\item \textbf{No high-coupling decoys}: removes high-coupling decoys but retains low-coupling decoys.
\squishend

\begin{table}[t]
\centering\small
\caption{Component ablation under the DeepSeek V4.1 Flash backbone and the avoidance-instruction condition, measured by decoy-page coverage. Each configuration contains 36 runs per scanner, and each value is computed over distinct URLs.}
\label{tab:rq2-ablation}
\setlength{\tabcolsep}{5pt}
\begin{tabular}{@{}lrrr@{}}
\toprule
Variant & AWE & HBGPT & YuraScanner \\
\midrule
Full \sys & 94.1\% & 92.5\% & 89.1\% \\
No low-coupling decoys & 92.4\% & 93.7\% & 9.0\% \\
No high-coupling decoys & 93.8\% & 89.2\% & 87.5\% \\
\bottomrule
\end{tabular}
\end{table}

\begin{figure}[t]
\centering
\includegraphics[width=\columnwidth]{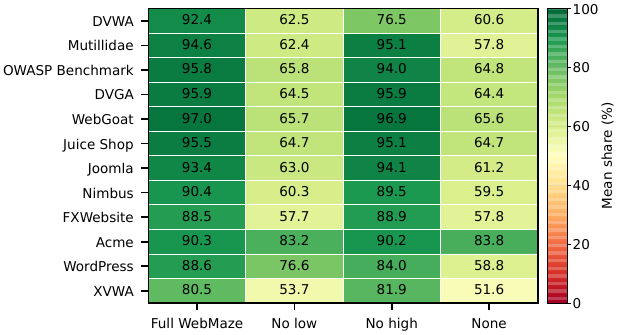}
\Description{Heat map with twelve application rows and four configuration columns, showing mean decoy-page coverage for the full configuration and three ablation variants.}
\caption{Per-application decoy-page coverage for the full configuration and ablation variants.}
\label{fig:rq2-ablation-site}
\end{figure}

As shown in \autoref{tab:rq2-ablation}, YuraScanner is affected differently by the two decoy classes.
Removing high-coupling decoys has little effect, reducing decoy-page coverage from 89.1\% to 87.5\%.
In contrast, removing low-coupling decoys reduces coverage to 9.0\%, and removing both classes further reduces it to 0\%.
The per-application results in \autoref{fig:rq2-ablation-site} show the same pattern.
These results show that YuraScanner relies mainly on low-coupling decoys, while high-coupling decoys provide additional entries into the decoy world.

Low- and high-coupling traps may affect legitimate users.
We therefore examine their impact on normal browsing by having one author manually use four versions of each application: without traps, with only low-coupling traps, with only high-coupling traps, and with the complete configuration.
The author observed no noticeable differences in normal browsing or application functionality across the four versions.
This is consistent with the integration mechanisms in \autoref{trap-inte}, which keep traps from interfering with genuine components and normal user interactions.

\subsection{RQ3: Scanner Awareness}
\label{sec:eval-awareness}
We evaluate whether \sys remains effective when scanners know about the decoy mechanism.
The decoy-page coverage experiment in RQ1 uses the no-disclosure
setting, in which scanners receive no additional information
about the deployed defense.
We additionally conduct two experiments that inform scanners about (1) the existence of the decoy world and (2) the existence of the decoy world together with instructions to avoid decoy pages.
Each experiment similarly contains 216 runs across scanners, applications, and model backbones.

\begin{table}[t]
\centering\small
\caption{Effect of scanner awareness on decoy-page coverage across the twelve-application campaign. Each value averages 36 runs. None denotes no disclosure, Aware denotes disclosure of the decoy world's existence, and Instructed additionally tells the scanner to explore only genuine functionality. Parentheses report the percentage-point change from None.}
\label{tab:rq3-awareness}
\resizebox{\columnwidth}{!}{%
\begin{tabular}{@{}lrrr|rrr@{}}
\toprule
& \multicolumn{3}{c|}{DeepSeek V4.1 Flash} & \multicolumn{3}{c}{GLM-5.3} \\
Scanner & None & Aware & Instructed & None & Aware & Instructed \\
\midrule
AWE & 94.3\% & 94.3\% (0.0) & 93.9\% (-0.4) & 94.3\% & 94.3\% (0.0) & 94.4\% (+0.1) \\
HBGPT & 97.3\% & 96.4\% (-0.9) & 94.8\% (-2.5) & 94.2\% & 92.9\% (-1.3) & 92.3\% (-1.9) \\
YuraScanner & 88.9\% & 89.4\% (+0.5) & 88.6\% (-0.3) & 88.9\% & 88.2\% (-0.7) & 85.6\% (-3.3) \\
\bottomrule
\end{tabular}
}
\end{table}

Disclosure of the decoy world's existence alone has little effect on decoy-page coverage.
As shown in \autoref{tab:rq3-awareness},
AWE remains at 94.3\% under both backbones, while HBGPT decreases by at most 1.3\% and YuraScanner changes by at most 0.7\% relative to no disclosure.
The largest reductions are 2.5\% points for HBGPT under DeepSeek V4.1 Flash and 3.3\% for YuraScanner under GLM-5.3.
Thus, even with explicit awareness of deception and avoidance instructions, coverage remains above 92\% for HBGPT and above 85\% for YuraScanner, and the scanners continue exploring the decoy world.
These results suggest that the scanners do not reliably distinguish genuine and decoy pages under the tested conditions.
Knowing that deception exists is insufficient to identify individual application-derived traps, which mimic legitimate web interfaces.

\subsection{RQ4: Decoy Deployment Statistics}
\label{sec:eval-footprint}

We characterize the deployment of \sys across 12 applications.
The original applications contain 1--200 genuine pages, while \sys generates 8--405 decoy pages per application.
This corresponds to 1.5$\times$--8.1$\times$ page expansion and 6.4$\times$--60.9$\times$ byte expansion.
Each deployment contains 43--7,183 injected placements and 228--237 registered decoy endpoints.

The amount of generated content and the number of injected placements vary substantially across applications, reflecting differences in their available views and placement opportunities.
In contrast, the number of registered decoy endpoints remains relatively stable.
Thus, \sys adapts the size of the decoy world and the number of entry points to each application without proportionally increasing the number of registered endpoints.
Detailed per-application statistics are reported in \autoref{tab:app-details}.

\section{Related Work}
\label{sec:related}
\PP{Web Scanning.}
Web vulnerability scanners interact with web applications through exposed interfaces and issue requests to explore application functionality and identify weaknesses~\cite{fong2007web}.
Prior tools augment this interaction by inferring application state machines, modeling client-side JavaScript, representing navigation graphs, or applying concolic execution~\cite{doupe2012enemy,pellegrino2015jak,eriksson2021blackwidow,alhuzali2018navex}.
Recent systems place LLMs in the exploration loop to interpret responses, plan actions, and adapt their behavior during testing~\cite{stafeev2025yura,happe2023hackingbuddygpt,jaswal2026awe}.
Other systems use LLMs primarily for planning and coordinate reconnaissance, retrieval, and exploitation components~\cite{deng2024pentestgpt,deng2026pentestgptv2,shen2025pentestagent}.
Although these systems differ in their internal analysis and planning mechanisms, they depend on application observations to guide subsequent exploration.
\sys targets this common observation--action loop and therefore does not depend on a particular scanner architecture.

\PP{Web Deception and Decoy Environments.}
Web deception techniques steer automated explorers toward misleading content or decoy environments instead of genuine application functionality.
Honeypot links and misleading page content offer destinations or instructions to automated explorers~\cite{wan2019pathmarker,mueter2008honeypots,ayzenshteyn2025cheat}.
Parallax deploys honeypot clones alongside live applications~\cite{tsouvalas2025parallax}, while WebCloak applies semantic interference against LLM-driven scraping~\cite{li2026webcloak}.
Automation fingerprinting instead identifies automated clients before they reason about application content~\cite{azad2020webrunner}.
AgentSnare constructs and validates artifacts based on an agent's interaction history to maintain a consistent decoy environment~\cite{wang2026agentsnarelearningdelaydivert}.
GenPot generates consistent responses for web interfaces, APIs, and web shells based on previous interactions~\cite{lara2026genpot}.
\sys complements these approaches by coupling deceptive entries to genuine application functionality and generating a view-tailored decoy world from the application's existing views.

\section{Conclusion}
\label{sec:conclusion}
We presented \sys, a web defense that combines functionality-coupled traps with a view-tailored decoy world to redirect LLM-based scanner exploration.
Across experiments on 12 applications, three scanners, and two model backbones, \sys directs an average of 93.0\% of the distinct requested pages into the decoy world.
This substantially limits scanners' access to vulnerability-related pages.
Even when scanners are informed about the defense and instructed to avoid decoy pages, their exploration behavior changes only slightly.
Together, these results establish functionality-coupled deception as a practical approach for redirecting autonomous web-scanner exploration while preserving genuine application functionality.
We believe functionality-coupled deception provides a promising foundation for protecting web applications as autonomous scanners become more capable.

\section*{Acknowledgments}
This work was supported in part by the National Natural Science Foundation of China (NSFC) under the Young Scientists Fund (Category C, Grant No. 62502355), and the NTU Start-up Grant.
Changhua Luo and Penghui Li are the corresponding authors.
\bibliographystyle{ACM-Reference-Format}
\bibliography{references}

\appendix
\section{Additional Evaluation Results}
\label{app:extended-evaluation}
This appendix collects supplementary evaluation results referenced in the main text.
It includes disclosure and backbone comparisons, component ablations, application statistics, and per-application backbone variation.

\subsection{Component Ablation}
\label{app:ablation}
We compare configurations that remove decoys injected into genuine pages while retaining the materialized decoy-world resources.
All configurations use the same target set and scanner-specific budget protocol.
For each site, we evaluate three configurations: \emph{Full \sys}, \emph{No low-coupling decoys}, and \emph{No high-coupling decoys}.
The removal groups follow the builder labels: high-coupling decoys include style binding, script splicing, and markup cloning, while low-coupling group includes independent traps.

Each configuration contains 108 runs: 12 sites $\times$ 3 scanners $\times$ 3 repetitions under disclosure with instruction.
AWE and YuraScanner use twenty minutes; HBGPT uses a hundred rounds.
Per-run contiguous traffic bursts define the measurement window, keeping reused proxy ports from extending a run.
\autoref{tab:app-ablation} reports decoy-page coverage for each scanner and configuration.

\begin{table}[t]
\centering
\caption{Component ablation under DeepSeek V4.1 Flash, avoidance-instruction condition. Each configuration has 108 runs (36 per scanner). Each value is mean decoy-page coverage, with min--max range in parentheses.}
\label{tab:app-ablation}
\resizebox{\columnwidth}{!}{%
\setlength{\tabcolsep}{2.5pt}
\begin{tabular}{@{}lrrr@{}}
\toprule
Variant & AWE & HBGPT & YuraScanner\\
\midrule
Full \sys & 94.1\% (78\%--100\%) & 92.5\% (82\%--98\%) & 89.1\% (75\%--97\%)\\
No low-coupling decoys & 92.4\% (70\%--100\%) & 93.7\% (84\%--100\%) & 9.0\% (0\%--75\%)\\
No high-coupling decoys & 93.8\% (77\%--100\%) & 89.2\% (11\%--99\%) & 87.5\% (66\%--94\%)\\
\bottomrule
\end{tabular}
}
\end{table}

\subsection{Application and Decoy-World Statistics}
\label{app:app-details}

\autoref{tab:app-details} summarizes the twelve applications and their generated decoy worlds.
For each application, we report the technology, numbers of genuine and decoy pages, injected placements, registered decoy endpoints, and the resulting page and byte expansion.

\begin{table}[t]
\centering\small
\caption{Application and decoy-world statistics. Pages reports genuine/decoy page counts; Pg.$\times$ and B.$\times$ report the decoy-to-genuine ratios in pages and bytes.}
\label{tab:app-details}

\begin{tabular}{@{}>{\raggedright\arraybackslash}p{.24\columnwidth}crrrrr@{}}
\toprule
Site & T & Pages & Pg.$\times$ & Inj. & EP & B.$\times$\\
\midrule
DVWA$^*$               & P & 20/40 & 2.0$\times$ & 727 & 230 & 16.1$\times$\\
Mutillidae$^*$         & P & 127/300 & 2.4$\times$ & 6{,}793 & 237 & 6.5$\times$\\
OWASP Benchmark$^*$    & J & 200/300 & 1.5$\times$ & 7{,}183 & 231 & 9.4$\times$\\
DVGA                   & G & 1/8 & 8.0$\times$ & 43 & 230 & 60.9$\times$\\
WebGoat                & J & 4/18 & 4.5$\times$ & 174 & 230 & 29.3$\times$\\
Juice Shop             & N & 87/240 & 2.8$\times$ & 3{,}718 & 231 & 18.6$\times$\\
Joomla                 & C & 5/40 & 8.0$\times$ & 277 & 228 & 18.0$\times$\\
Nimbus                 & H & 8/40 & 5.0$\times$ & 491 & 230 & 12.9$\times$\\
FXWebsite              & H & 27/162 & 6.0$\times$ & 1{,}578 & 231 & 12.0$\times$\\
Acme                   & F & 12/35 & 2.9$\times$ & 375 & 230 & 22.1$\times$\\
WordPress              & C & 50/405 & 8.1$\times$ & 3{,}991 & 236 & 6.4$\times$\\
XVWA                   & P & 26/162 & 6.2$\times$ & 1{,}500 & 236 & 10.1$\times$\\
\bottomrule
\end{tabular}

\par\smallskip\scriptsize
$^*$ Applications used for the vulnerability-page coverage evaluation.
Inj.: injected placements; EP: registered decoy endpoints.
T denotes technology: F, Python/Flask snapshot; G, Python/GraphQL;
P, PHP/MySQL; H, static HTML; C, PHP/MySQL CMS;
N, Node.js/Angular; J, Java/Spring.
\end{table}

\subsection{Backbone Sensitivity}
\label{app:model-sweep}
\begin{figure}[t]
\centering
\includegraphics[width=\columnwidth]{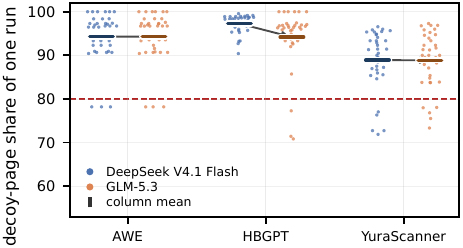}
\Description{One panel with the three scanners along the x axis and the
decoy-page share of a single run on the y axis. Each scanner carries two columns
of dots, one dot per run, blue for DeepSeek V4.1 Flash and orange for GLM-5.3, a short bar
at each column's mean and an arrow from one mean to the other. The columns
overlap over most of their range; the dashed line marks 80 percent.}
\caption{Runs from the two backbone batches, one dot per run, with the
twelve applications pooled into the scanner they were run against: blue dots are
DeepSeek V4.1 Flash runs and orange dots GLM-5.3 runs, the bar is the mean of its column
and the arrow goes from one mean to the other. The dashed line marks 80\%;
\autoref{tab:app-model-site} gives per application comparison.}
\label{fig:app-model-sweep}
\end{figure}

The main text reports aggregate disclosure and backbone results in \autoref{tab:rq3-awareness}.
This subsection provides the corresponding run-level distribution and per-application differences.
\autoref{fig:app-model-sweep} shows the run distribution and \autoref{tab:app-model-site} reports the per-application differences between the two backbones.
\begin{table}[t]
\caption{The per-application change behind \autoref{fig:app-model-sweep}, in the same
application order: GLM-5.3 minus DeepSeek V4.1 Flash, in points of mean decoy-page share,
with nine runs per scanner and application. The last column pools the three
scanners.}
\label{tab:app-model-site}
\centering\small
\begin{tabular}{@{}l rrrr@{}}
\toprule
Application & AWE & HBGPT & Yura & All \\
\midrule
DVWA$^*$               & +0.0 & -0.9 & -4.1 & -1.6 \\
Mutillidae$^*$         & +0.2 & -1.7 & -1.1 & -0.9 \\
OWASP Benchmark$^*$    & +0.0 & +2.2 & -4.7 & -0.9 \\
DVGA                   & +0.0 & -3.2 & -0.8 & -1.3 \\
WebGoat                & +0.0 & +0.9 & -1.5 & -0.2 \\
Juice Shop             & +0.0 & -2.1 & +0.3 & -0.6 \\
Joomla                 & +0.5 & -3.3 & +0.0 & -1.0 \\
Nimbus                 & +0.0 & -1.4 & -3.2 & -1.5 \\
FXWebsite              & +1.3 & -2.8 & +0.7 & -0.3 \\
Acme                   & +0.0 & -18.7 & -1.4 & -6.7 \\
WordPress              & +0.2 & +1.1 & -5.7 & -1.5 \\
XVWA                   & +0.0 & -6.3 & +4.5 & -0.6 \\
\bottomrule
\end{tabular}
\end{table}

\PP{Per-Application Variation}
AWE's mean changes by at most 1.3 points on every application. Yura changes
by more than 2 points on five applications, including WordPress
(77.5\% to 71.8\%), OWASP and DVWA. HackingBuddyGPT changes by more than
2 points on seven applications and by more than 4 on XVWA
(92.9\% to 86.6\%) and Acme (93.8\% to 75.1\%). Its Acme means are
6.3 genuine and 19.7 decoy pages under GLM-5.3, versus 8.7 and 149.7 under
DeepSeek V4.1 Flash; XVWA has 3.9 genuine and 11.0 decoy pages under GLM-5.3. These
counts expose the denominator underlying the share changes. The lowest
per-application mean remains 71.8\%, on WordPress under GLM-5.3.
The two backbones thus retain substantial decoy concentration despite their
different traffic volumes and application-level exploration patterns.

\end{document}